# THE DECISION CRITERIA USED BY LARGE ENTERPRISES IN SOUTH AFRICA FOR THE ADOPTION OF CLOUD COMPUTING

Tserwa Bakasa, University of Cape Town, bskste001@myuct.ac.za

Ayanda Pekane, University of Cape Town, ayanda.pekane@uct.ac.za

**Abstract:** Cloud computing is a technology that has become increasingly popular over the past decade within several enterprises. This popularity can be attributed to its benefits, including lower operating costs, improved computational capabilities, increased flexibility and on-demand storage space. As a result, many enterprises are already in various Cloud Computing (CC) adoption and implementation stages. This study investigates the decision criteria used by large enterprises in South Africa (SA) for the adoption of cloud technology. The majority of large enterprises have comprehensive resources, resulting in established Information Technology (IT) systems and infrastructure set up within their organizations. Though this is the case, the adoption of CC by large enterprises has been on the rise. This may not be a surprise as CC literature points out to benefits and influencers of CC adoption. However, the decision criteria used by large enterprises in SA in adopting CC are lacking in the literature reviewed. The study followed an inductive approach making use of qualitative methods. Findings revealed that large enterprises do not make use of a formalized or standardized decision criteria. However, operational cost, enterprise strategic intent and product efficiency formed key criteria for adopting CC. In addition, security, cloud service provider adoption frameworks and data sovereignty were the key criteria used to select a CC service provider. The research will contribute towards CC technology adoption literature, particularly for developing countries.

## 1. INTRODUCTION

Cloud computing is a technology that has become increasingly popular within large enterprises in SA. The reason behind this popularity can be attributed to the multiple benefits that are offered by cloud technology. Sharma and Sehrawat (2020) state that the most outstanding benefits of CC include lower expenses, increased flexibility, scalability, green technology, increased computing power, remote access and innovation. Overall, this equates to lower overhead costs, improved computational and information-handling capabilities, and greater convenience. These advantages make CC an attractive option to consider for enterprises in general (Ross, 2010). The significance of large enterprises in SA is evident in that they employ the highest South African populous (StatsSA, 2019). The small business institute of South Africa reports that small and medium enterprises use only 28% of the South African population.

In comparison, 72% of the population is employed by large enterprises (Smallbusinessinstitute, 2018). Additionally, large enterprises are the highest contributors to SA's government revenue stream. According to StatsSA, the formal business sector realized a total turnover of approximately R2,39 trillion in 2019. Out of this amount, large enterprises contributed the highest percentage of 62%, while small and medium-sized businesses contributed 29% and 10%, respectively (StatsSA, 2019). These statistics clearly illustrate the critical role that is played by large enterprises in the South African economy.

Large enterprises are also characterized by possessing comprehensive resources resulting in established Information Technology (IT) systems and infrastructure set up within their organizations. Though this is the case, the adoption of CC by large enterprises has been on the rise. This may not be surprising, as the literature points out the benefits of CC adoption. The reviewed literature further outlines additional themes such as the challenges, factors and influences of CC adoption (Johnston, Loot & Esterhuyse, 2016; van Dyk & Van Belle, 2019). These themes have advanced the body of knowledge in the adoption and use of IT in SA. However, the decision criteria used by large enterprises in SA in adopting CC are lacking in the literature reviewed. Mar (2018) defines the decision criteria as a set of guidelines or principles that are used to make a decision. The guidelines and principles make up specifications that are measurable and should amount to some form of a scoring system. According to Ross (2010, p. 9), "the decision on whether or not to adopt cloud computing can be a difficult decision for managers to make, considerable thought goes into making such a decision." However, the aspect of the decision criteria involved in the adoption of CC has not been explored in the context of enterprises of SA. The objective of this study is to describe the decision criteria used by large enterprises in SA to adopt CC. The study seeks to answer the following research question; What are the decision criteria used by large enterprises in SA to adopt CC?

This study aims at contributing to the current body of literature by promoting an understanding of the decision criteria used by large enterprises to adopt CC in SA. In addition, the study aims to identify the decision criteria used to adopt CC in the South African context without adopting theories developed in the global North. The research will be relevant to the community of practice interested in the decision criteria used in adopting CC.

## 2. LITERATURE REVIEW

The literature reviewed for this study includes the adoption of CC literature in developing countries, including SA, the decision-making influences in adopting CC as a technology, and enterprises' decision criteria in adopting technologies. Literature on decision making was included due to the close link between decision making and decision criteria.

### 2.1 The adoption of CC in large enterprises in SA

Cloud computing is a common pool of configurable IT services, including data storage, networks, information processing and software applications rendered as a flexible and scalable service via the internet (Akande, 2014). According to Rahimli (2013), CC is suitable for large enterprises like financial services, government organizations and healthcare facilities. This is because CC provides increased data storage, minimum downtime and improved processing ability to facilitate quicker responses and more reliable security (Rahimli, 2013; Sharma & Sehrawat, 2020). Therefore, the adoption of CC results in lower expenses, increased flexibility, scalability, green technology, increased computing power, remote access and innovation (Sharma & Sehrawat, 2020). This has made CC an attractive option to consider for large enterprises (Ross, 2010). As a result, large enterprises in South Africa have started to gravitate towards cloud adoption to utilize emerging technologies, reduce costs and improve efficiency (Moonasar & Naicker, 2020).

Oliveira, Thomas and Espadanal (2014) elaborates, that large businesses have comprehensive resources, they have maintained and upgraded their IT infrastructure. They can also move on from the traditional way of using in-house IT infrastructure to introducing CC in their enterprise (ibid.). The literature reviewed revealed themes that are associated with the adoption of CC by large enterprises. However, the literature reviewed also showed that publications on CC adoption by large enterprises in SA are under-represented.

The literature on CC adoption in SA by large enterprises is concentrated on descriptive and explanatory contributions. The themes include the level of readiness to adopt CC (Akande, 2014); the business value of CC in organizations (Johnston et al., 2016); the operational requirements for the adoption of CC (Krauss & Van der Schyff , 2014); barriers to the adoption CC (Scholtz et al., 2016). The contributions progress to prescriptive works consisting of a framework for CC adoption with a focus on complying with the Protection of Personal Information Act (POPI Act) (Skolmen & Gerber, 2015). Carroll, Van Der Merwe and Kotze (2011) focused on mitigating CC security risks. One of the relevant themes found in literature was the decision-making factors that influence CC (Alshamaila, Papagiannidis, & Li, 2013).

## 2.2 Decision-making factors that influence CC adoption in large enterprises in SA

Research shows that several factors can influence decision-makers' decisions to adopt CC within their respective organizations. These factors consist of competitive market pressures (Alshamaila et al., 2013; van Dyk & Van Belle, 2019), Institutional pressures (Liang, Saraf, Hu, & Xue, 2007; Trope, 2014), Senior management support (Priyadarshinee, Raut, Jha, & Garda , 2017; Sharma, Gupta, & Acharya, 2020), A need for improved IT services (Priyadarshinee et al., 2017) and core business focus (Awosan, 2014). Given all the factors influencing decision-makers to adopt CC, the literature proposes different CC adoption decision criteria frameworks. Most of these frameworks are based on the global north perspective.

## 2.3 Cloud computing adoption decision criteria and frameworks

Alhammadi (2016) presents a decision-making framework known as the Knowledge Management Based Cloud Computing Adoption Decision Making Framework (KCADF). With this framework, the decision-making process is categorized into three explicit levels. These levels are strategic, operational and tactical decision levels. The main decision as to whether an enterprise should adopt CC or not, is made during the strategic decision level stage. The next decision-making step is the tactical decision-making phase. This decision point involves the decision to select a cloud deployment model. The third and final decision stage is the operational decision level. This level involves deciding on the type of cloud service module to adopt.

Alhammadi (2016), further proposes the analytic hierarchy process (AHP) as a decision criteria tool. This tool aims to support the KCADF. By definition, AHP is "a multi-criteria decision-making approach in which factors are arranged in a hierarchic structure" (Saaty, 1990, p. 9). This decision model has three central pillars: the hierarchical structure of the model; the pairwise comparison of the available options and criteria; and the fusion of priorities. Saaty (1990) states that the problem-solving objective takes priority of the hierarchy. This is followed by the decision-making criteria and lastly, the alternative solutions.

As reported by Kaisler, Money, and Cohen (2012), CC adoption decisions are often made without a stringent analysis. Kaisler et al. (2012) proposed a three-tier decision framework model. These decision levels are the service, system and application architectural decisions. The authors argue that the first step for decision-makers is to ascertain if a CC solution architecture is compatible with the enterprise's current applications. The next step consists of determining the level of adaptability between the solution's architecture and the enterprise's application. The last step is the application architecture stage associated with three decision areas: partitioning, scaling, and integration when it comes to the capabilities of the enterprise.

## 3. RESEARCH METHODOLOGY

This study adopted an inductive approach with an interpretive perspective. A qualitative methodology was followed using a case study method (Yin, 2018, van Dyk & Van Belle, 2019). The target population for this research was made up of the following; IT infrastructure managers, enterprise and solution architects working in large enterprises in various industries. This target population was selected because it is the population that is primarily responsible for making the decisions that determine an enterprise's CC adoption strategy. The purposive sampling technique and snowballing were adopted (Etikan, Alkassi & Abubakar, 2016; Yadav, Singh & Gupta, 2019).

This research project used unstructured interviews as a data collection technique. This was the most appropriate data collection technique to use since the research project followed an inductive approach. According to Woo, O'Boyle and Spector (2017), an inductive approach is not bound by specific hypotheses or theory. This is because the most imperative attribute of inductive research is the identification of a new phenomenon. By following an inductive approach, the researcher was able to identify new phenomena from the data without being bound to any pre-existing hypotheses or theories. Unstructured interviews allowed the respondents to give experiences that were unbiased by any perspectives or past research findings. While the researcher had conducted a literature review, they were careful not to impose their views or previous experience on the respondents.

The interviews were conducted and recorded on the Microsoft Teams online platform. This was because of the COVID 19 pandemic, which resulted in travel restrictions and social distancing regulations. The sample size for this research comprised of a total of eight (8) participants. However, only seven (7) participants were available for interviews. The average duration of the interviews ranged between 45- 60 minutes. This small sample size was appropriate for this research because the study was qualitative. Creswell and Guetterman (2019) note that qualitative research studies investigate a few individuals to understand the central phenomenon that is being studied. In addition, a cross-sectional time frame design was applied for this research hence the small sample. A cross-sectional study was the most appropriate timeframe to use as the project was carried out over a short timeframe. The research used the cross-sectional timeframe to capture a snapshot of the characteristics of the research phenomenon over the few months of the project's duration. The respondents, their roles, industries, and motivation for selecting respondents are summarized in Table 1.

| Enterprises | Industry | Respondent Position | Reason for respondent selection |
| --- | --- | --- | --- |
| Enterprise A | Government department (Research Institute) | Senior Enterprise Architect | The respondent was a member of the cloud adoption decision making team. |
| Enterprise B | Mobile telecommunications | Enterprise Architect Lead | The respondent was the team lead of the cloud adoption team. |
| Enterprise C | Information Technology (System Integration) | Senior Practice Lead | The respondent was directly involved in the cloud adoption decision process. |
| Enterprise D | Private healthcare | Enterprise Architect Lead | The respondent was the team lead of the cloud adoption decision making team. |
| Enterprise E | Energy & chemical | Data Architect/Warehouse Lead | The respondent was part of the cloud migration decision making team. |
| Enterprise F | Investment and Insurance | Business Strategy Delivery Lead | The respondent formed part of the cloud migration decision making team. |
| Enterprise G | Banking | Solution Architect | The respondent was part of the cloud migration decision making team. |

**Table 1 Breakdown of the respondents**

The researcher followed Braun and Clarke's (2012) six-phase guide to analyze collected data using a thematic analysis approach. The researcher familiarized themselves with the collected data. This was achieved by listening to the audio recordings which were recorded for the interview. The recordings were transcribed, read and notes were made during this process to gain a deeper understanding of the data collected. Initial codes were then generated by identifying common patterns within data (Nvivo was used as the software data analysis tool). The researcher continues the analysis process by searching for themes. According to Braun and Clarke (2012), the creation of the themes needs to be guided by the study's research question. Therefore, themes were created based on the research question of the study: What are the decision criteria used by large enterprises in South Africa to adopt cloud computing? The themes were then grouped into two categories, namely; the cloud adoption decision criteria and the cloud service provider decision criteria.

The themes were reviewed and carried out with the coded datasets. This phase amounted to a quality analysis procedure as it checked the created themes against the existing data. The next step involved defining and naming the themes. Lastly, the findings are presented and discussed. The next sections describe the category of themes that emerged from the process of data analysis.

## 4. FINDINGS

The findings revealed that large enterprises in SA do not use a standardized criteria to adopt CC. Table 2 outlines themes and quotations from respondents.

| THEME IDENTIFIED | |
|---|---|
| 1. Operational costs | *"And that was our biggest frustration is the data centers. So, we have 52 data centers around the country. But there are inherent costs in expanding data center space. That's where we wanted to expand that one floor in Cape Town was going to be needed nearly 100 million Rand just to do the infrastructure piece, not even the hardware that goes in, but to do the conditioning, and the flooring, and the fires and stuff."* RP02 |
| 2. Enterprise strategic intent | *"We knew we needed to go to the cloud, we knew that we were trying to achieve a strategy, which is going digital, we needed a flexible environment, we were in trouble with stability of our data centers. We needed a framework that enables us to get to the digital enterprise, the digital vision."* RP04 |
| 3. Product efficiency | *"So, when you're investing in an infrastructure platform, what's next? What're the additional services that are provided? So, stuff on a basic level, for example, with Azure, you get SQL on Azure on a platform layer, you can start integrating with, for example, IoT services, or whatever the case might be."* RP04 |
| 4. Security | *"Then outside of that the kind of the broader topics that we had to resolve, there were big security concerns, we had to get at risk, corporate risks to sign off on this, we had to put down very detailed security policy around cloud use."* RP03 |
| 5. Previous industry case studies | *"We had to look at a lot of industry case studies. And so basically, because they're an insurance company, we had to look at how other insurance companies have implemented the strategy, how it worked for them"* RP06 |
| 6. Cloud service provider adoption framework | *"We looked Amazon. So, Amazon has got a great cloud adoption framework. That's one of the frameworks that we considered"* RP01; *"Because, geez, so what we've done so far, we've set out, okay, we said let's look at the different providers that are there. What frameworks do they have in terms of informing how their services are consumed?"* RP03 |
| 7. Storage scalability | *"The ability to obviously, provide the scalable infrastructure that's immediately available to people, especially if you've got a system integration environment whereby you needed to spin up development environments quite quickly for developers…."* RP04 |
| 8. System integration compatibility | *"Because from an application point of view, we have to consider the integration requirement"* RP01; *"Yeah, obviously, does our current infrastructure support the move to cloud? Will the systems be able to communicate well in accessing the data? Reading the data, translating the data?"* RP06 |
| 9. Data sovereignty | *"You're going to need to justify when you look into the guise of like, know, the governance and the regulatory, you need to prove to them that whatever that you thought you'd be doing will be not violating their principles or, or their regulations like the POPI, kind of like regulations must be met."* RP04 |
| 10. Time to build | *"It's a four-week process of procuring hardware, plus another two weeks installing or four weeks installing the software, cloud is immediately available once you've got that synchronization done."* RP03 |
| 11. Data center location | *"…the reason why we ended up implementing the cloud is that Amazon was, was about to build a data station in Cape Town. Yes, and played a very big factor because you don't want your data sitting offshore."* RP06 |
| 12. Data latency | *"And then the second thing was latency. So obviously, we have performance and latency dependent applications. And it's about 130 milliseconds latency from South Africa to Europe, which can have a detrimental effect."* RP02; RP04: *"So if you want if you're in a hospital, and you have to go to Google in America, or Ireland, Europe, or it could be Australia or Brazil, that hop across the ocean is going to cause a delay. So that was a very, very big deciding factor against Google."* RP04 |
| 13. Latest technology trends | *"What we're looking for this is this is the latest trends in terms of the different applications that are there and so on and so on."* RP01; *"And being a leader in the industry is obviously important for us as an IT organization, you need to show it in our technology. We call it you eat your own dog food. You know, you need to know, or somebody else calls or drink your own champagne, but you need to live and understand technologies and being able to showcase it".* RP02 |
| 14. Disaster Recovery strategy | *"What happens if the cloud shuts down? Do we have backup? Is it restored onto a different cloud platform? is it? Yeah, there were all of those considerations that we had to take into account."* RP06 |

**Table 2: Decision criteria themes**

The finding revealed that the decision criteria are divided into two main categories. Namely, the cloud technology adoption decision criteria and the cloud service provider decision criteria. The cloud technology adoption category refers to the decision criteria used for the adoption of CC. In this category, decision-makers are considering the actual adoption of CC. The themes consist of operational cost, enterprise strategic intent, product efficiency, previous industry case studies, scalability, system integration compatibility, time to build and latest technology trends.

### 4.1 Cloud computing adoption decision criteria

Operational cost is one of the key determining criteria considered in the cloud adoption decision by large enterprises in SA. This is where a comparison is often made between the operational costs of traditional data centers and CC. Sharma and Sehrawat (2020) note that CC adoption results in a significant reduction of operational costs within an enterprise. An enterprise's strategic intent is another criterion that large enterprises use. The findings revealed that an enterprise's IT strategy would shape an enterprise's decision on whether or not to adopt cloud technology. Findings have also revealed that the efficiency of the cloud product compared to other alternative solutions is a fundamental criterion used during the cloud adoption decision process. These alternative solutions include in-house data centers and applications. Cloud technology comes with a bundle of additional services in comparison to other alternative solutions.

Large enterprises used previous industry case studies as a cloud adoption decision criterion. The case studies consist of enterprises that have similar challenges that operate in the same industry. Previous case solutions would be analyzed, focusing on how CC was used to address challenges. The next criterion is the scalability of storage space. Cloud solution offers more a scalable product which allows an enterprise to increase its storage capacity at any time (Techopedia, 2012). System integration compatibility is a criterion that involves the investigation of the current systems that exist within the enterprise to see whether they would be compatible with the cloud solution integration.

Another criterion is the concept of time to build. This refers to the lengthy time it would take to build up or build new IT applications and infrastructure instead of adopting CC. The last criterion refers to the need to keep up with the latest technological trends. This involves keeping up with the latest software applications, analysis tools and business delivery and applications. This is particularly important for the requirement of IT flexibility in enterprises. The next level of decision-making is the selection of a CC service provider.

### 4.2 Cloud computing service provider decision criteria

The cloud service provider decision criteria refer to decision criteria used to select the cloud service provider. The findings revealed security, cloud service provider adoption framework, data sovereignty, data latency, data centre location and the disaster recovery strategy as the decision criteria in this category.

Security is one of the key factors considered for the cloud service provider selection criteria by large enterprises in SA. Kaufman (2009) reported that security plays an important role in the adoption of CC. The next criterion in the category is cloud adoption frameworks that major cloud service providers provide to possible users. The cloud adoption frameworks consist of documents that include literature, with cloud adoption steps involved when using a particular cloud service provider. The frameworks outline cloud services that the cloud service provider offers.

Data sovereignty can be described as a government's control and authority and overall the data stored within that country (Irion, 2012). It was one of the criteria for selecting the cloud service provider. Respondents cited concerns with governance issues like the fact that they need to adhere to regulatory requirements specific to South Africa, like the POPI act. Data latency was another consideration, Weissberger, McWhirter, Klassen, and Ginsberg (2004) describe data latency as a time delay in the movement of data packets from one point to the next. Depending on the industry, data latency can be a deal-breaker for example; a respondent made an interesting point that "*…if you're in a hospital, and you have to go (connect) to Google in America, or in Ireland, Europe…across the ocean, it's going to cause a delay. So that was a very, very big deciding factor..."* RP04.

Data center location is one of the criteria that large enterprises are considering. Enterprises in SA prefer a cloud service that has local data centers. This is in comparison to cloud service providers that have data centers that are based overseas. The reason for this can be attributed to data privacy and data latency concerns. The last criterion in this category is disaster recovery strategy. A disaster recovery strategy can be defined as a plan to effectively restore IT operations after a service disruption. Several factors have involved this strategy. These factors include the acceptable outage time, the security aspect, the cost of the recovery plan and the system integration logistics (Fallara, 2004).

## 5. DISCUSSION

This study aims to describe the decision criteria used by large enterprises in SA when adopting CC. A decision criterion is a set of guidelines that make up measurable specifications that often amount to a scoring system (Mar, 2018). This research revealed that large enterprises do not only decide whether to adopt CC but also consider selecting the appropriate CC service provider. The findings further revealed that large enterprises don't use standardized or formalized cloud adoption decision criteria when adopting CC. Though there aren't any standardized decision criteria used, formal processes, appointed teams, and individuals form part of the decision-making. The decision criteria fundamentally include operational cost, enterprise strategic intent, product efficiency, previous industry case studies and scalability for CC adoption. This is in line with the literature that was reviewed on the benefits that CC affords larger enterprises (Johnston et al., 2016; Sharma & Sehrawat, 2020). Additionally, Alhammadi's (2016) KDCAF cloud adoption framework resembles similarities with the CC adoption criteria findings of this study.

The first criterion that is included in the KDCAF framework is the initial strategic decision to adopt CC. Similarly, the findings reveal that one of the key criteria that large enterprises consider when adopting CC is the need to align with their strategic intent. The second and third levels in the KDCAF framework are the operational and tactical decision levels. These levels would include the cloud deployment model and service models, respectively. The decision-making at these levels focuses on on-demand service, flexibility, operational costs, and product quality criteria. The findings of this research also reveal that time to build infrastructure, operational costs, and scalability are the main criteria for adopting CC by large enterprises in SA.

The main difference between the KDCAF cloud adoption framework and the criteria that are revealed in this study is the decision-making level of selecting CC service providers. Findings of this research further revealed that security, cloud service provider adoption framework, data sovereignty, data latency, data center location and the disaster recovery strategy are considered for selecting a preferred CC service provider. Previous literature on the adoption of CC in SA does not expand on enterprises' selection of CC service providers. However, Priyadarshinee et al., (2017) emphasize that each enterprise is different and will have its unique set of variables on a decision framework. In conclusion, the inductive approach that the study followed allowed for these decision criteria to emerge from data without imposing a theoretical framework, particularly from the global north perspective. The criteria used for selecting CC service providers reveal some contextual matters as underlying reasons for this criteria. For example, at the heart of data sovereignty is the consideration of national security matters which have implications on compliance with South African regulations (Irion, 2012).

Another example is that of data latency and data center location. Enterprises in SA prefer a cloud service that has local data centers. This is because data centers that are based overseas have a time delay in the movement of data packets from one point to the next, this may impact efficiency and productivity in enterprises (Weissberger et al., 2004). This research only focuses on large enterprises in South Africa. This was due to the time constraint of the research project. Future research could broaden the research scope and investigate large enterprises in other developing countries outside of the South African borders. Researchers may also compare the decision criteria used in South African enterprises and those used in other developing economies.